\documentclass[notitlepage,nofootinbib,preprintnumbers,amssymb,superscriptaddress]{revtex4-1}

\usepackage{amsfonts,amssymb,amsmath,graphicx,color,bm}
\definecolor{ultramarine}{rgb}{0.07, 0.04, 0.56}
\definecolor{cadmiumgreen}{rgb}{0.0, 0.42, 0.24}
\definecolor{indigo(dye)}{rgb}{0.0, 0.25, 0.42}
\usepackage[dvipdfmx]{hyperref}
\hypersetup{
colorlinks=true,
citecolor=ultramarine,
linkcolor=cadmiumgreen,
urlcolor=indigo(dye),
}

\newcommand{\fr}[2]{\frac{#1}{#2}}
\newcommand{\pa}{\partial}
\newcommand{\ti}{\tilde}
\newcommand{\na}{\nabla}

\newcommand{\del}{\delta}
\newcommand{\pb}[1]{\brc{#1}_{\rm P}}
\newcommand{\bra}[1]{\left( #1 \right)}
\newcommand{\brb}[1]{\left[ #1 \right]}
\newcommand{\brc}[1]{\left\{ #1 \right\}}
\newcommand{\be}{\begin{equation}}
\newcommand{\ee}{\end{equation}}
\newcommand{\bem}{\begin{bmatrix}}
\newcommand{\eem}{\end{bmatrix}}
\newcommand{\al}{\alpha}
\newcommand{\ga}{\gamma}
\newcommand{\Ga}{\Gamma}

\newcommand{\la}{\lambda}
\newcommand{\si}{\sigma}
\newcommand{\vp}{\varphi}
\newcommand{\Om}{\Omega}
\newcommand{\mn}{{\mu \nu}}
\newcommand{\mA}{\mathcal{A}}
\newcommand{\mB}{\mathcal{B}}
\newcommand{\mC}{\mathcal{C}}
\newcommand{\mE}{\mathcal{E}}
\newcommand{\mF}{\mathcal{F}}
\newcommand{\mG}{\mathcal{G}}
\newcommand{\mH}{\mathcal{H}}
\newcommand{\mK}{\mathcal{K}}
\newcommand{\mL}{\mathcal{L}}
\newcommand{\mM}{\mathcal{M}}

\newcommand{\mQ}{\mathcal{Q}}
\newcommand{\mR}{\mathcal{R}}
\newcommand{\mU}{\mathcal{U}}
\newcommand{\mV}{\mathcal{V}}

\newcommand{\sizecorr}[1]{\makebox[0cm]{\phantom{$\displaystyle #1$}}}
\newcommand{\sqx}{\sqrt{2X}}

\begin{document}

\preprint{RUP-18-30}

\title{Extended Cuscuton:~Formulation}

\author{Aya Iyonaga}
\affiliation{Department of Physics, Rikkyo University, Toshima, Tokyo 171-8501, Japan}

\author{Kazufumi Takahashi}
\affiliation{Department of Physics, Rikkyo University, Toshima, Tokyo 171-8501, Japan}

\author{Tsutomu Kobayashi}
\affiliation{Department of Physics, Rikkyo University, Toshima, Tokyo 171-8501, Japan}

\begin{abstract}
Among single-field scalar-tensor theories, there is a special class called ``cuscuton,"
which is represented as some limiting case of k-essence in general relativity.
This theory has a remarkable feature that the number of propagating degrees of freedom is only two in the unitary gauge in contrast to ordinary scalar-tensor theories with three degrees of freedom.
We specify a general class of theories with the same property as the cuscuton
in the context of the beyond Horndeski theory,
which we dub as the extended cuscuton.
We also study cosmological perturbations
in the presence of matter
in these extended cuscuton theories.
\end{abstract}

\maketitle

\section{Introduction}\label{sec:intro}
The framework of scalar-tensor theories has been extensively studied as a simple
and interesting
extension of general relativity (GR) and innumerably many models have been proposed.
These theories have been employed as a powerful tool to study the late-time cosmic acceleration and/or inflation in the early universe.
To handle such diverse models efficiently, some unifying descriptions of scalar-tensor theories have been developed so far.
The well-known example is the Horndeski theory~\cite{Horndeski:1974wa,Deffayet:2011gz,Kobayashi:2011nu}, which is the most general single-field scalar-tensor theory in four dimensions whose Euler-Lagrange equations are at most of second order.
This nature is desirable as it offers a unique class of scalar-tensor theories that can trivially avoid unstable extra degrees of freedom (DOFs) associated with higher-order equations of motion (EOMs), namely Ostrogradsky ghosts~\cite{Woodard:2015zca}.
One should note that the Horndeski theory is {\it not} the most general class that is free of Ostrogradsky ghosts: In the Lagrangian formalism, the Ostrogradsky instability can be circumvented if the system of Euler-Lagrange equations is degenerate and hence the higher derivative terms
can be eliminated by taking linear combinations of the EOMs.
Equivalently, in the Hamiltonian language, an additional primary constraint arises due to the degeneracy, which eliminates the problematic Ostrogradsky ghost.
There have been some attempts to construct scalar-tensor theories that surpass the Horndeski class~\cite{Motohashi:2016ftl,Crisostomi:2017aim}, and some broader classes
without Ostrogradsky ghosts have been discovered, including the Gleyzes-Langlois-Piazza-Vernizzi
(GLPV, also known as beyond Horndeski)
theory~\cite{Gleyzes:2014dya} and degenerate higher-order scalar-tensor
(DHOST, also known as extended scalar-tensor)
theories~\cite{Langlois:2015cwa,Crisostomi:2016czh,BenAchour:2016fzp}.

Generically, the above scalar-tensor theories without Ostrogradsky ghost have three DOFs, which consist of two tensor modes and one scalar mode.
However, there is a special class called ``cuscuton"~\cite{Afshordi:2006ad}, in which only two DOFs propagate and the scalar mode is nondynamical in the unitary
gauge, $\phi=\phi(t)$~\cite{Gomes:2017tzd}. 
The action of the cuscuton theory is written as
    \be
    S=\int d^{4}x\sqrt{-g}\brb{\fr{\mR}{2\kappa^2} + \mu^{2}\sqrt{2|X|} - V(\phi)}, \label{originalcuscuton}
    \ee
where $\mR$ is the four-dimensional Ricci scalar, $X\equiv -g^{\mn}\pa_{\mu}\phi\pa_{\nu}\phi/2$, and $\kappa$ and $\mu$ are nonvanishing constants.
It should be noted that the cuscuton is the unique\footnote{In fact,
$\mu$ may be a function of $\phi$,
but one can always make a field redefinition so that $\mu$ is just a constant.}
subclass of
the k-essence theory (i.e., GR plus ``$P(\phi,X)$''~\cite{ArmendarizPicon:1999rj})
respecting the two-DOF nature.
This in particular implies
the following properties:
\begin{enumerate}
\renewcommand{\theenumi}{\Alph{enumi}}
\renewcommand{\labelenumi}{[\theenumi]}
\item \label{charA} \textit{The field equation of the scalar field is at most of first order in the case of homogeneous and isotropic cosmology.}
In this cosmological setup,
one may safely choose the unitary gauge~$\phi=\phi(t)$. Then, the second term in Eq.~\eqref{originalcuscuton} takes the form $\mu^{2}|\dot{\phi}|$ with a dot denoting $\pa /\pa t$, so the Euler-Lagrange equation for $\phi$ does not contain second or higher derivatives of $\phi$.
Thus, the scalar field becomes nondynamical and its evolution is determined
by the dynamics of the metric through the constraint equation.
\item \label{charB} \textit{The kinetic term of scalar cosmological perturbations vanishes.}
If the action~\eqref{originalcuscuton} is expanded to second order in scalar perturbations
around a cosmological background,
one ends up with the quadratic action for a single variable~$\zeta$ (the curvature perturbation), where it turns out that the coefficient of the kinetic term $\dot{\zeta}^{2}$ vanishes.
This is due to the nondynamical nature of $\phi$ in the cuscuton theory.
\end{enumerate}
As is anticipated,
the two properties~[\ref{charA}] and [\ref{charB}] are closely
related to each other (see \S \ref{sec:formulation}).

Various aspects of the cuscuton
make this model fascinating.
Although it has the same number of physical DOFs as GR, the cuscuton theory exhibits some peculiar features, e.g., in the cosmic microwave background and matter power spectra, which can be distinguished from GR~\cite{Afshordi:2007yx}.
The authors of Ref.~\cite{deRham:2016ged} showed the absence of caustic singularities
in cuscuton-like scalar-field theories.
It was also pointed out that the cuscuton theory with a quadratic potential is
considered as a low-energy limit of the (non-projectable) Ho\v{r}ava-Lifshitz theory~\cite{Afshordi:2009tt,Bhattacharyya:2016mah}.
Stable bounce cosmology based on the cuscuton
has been studied in Ref.~\cite{Boruah:2018pvq}.

Given such advantages, it would be intriguing to find more general
theories which share the same nature as the cuscuton model, i.e.,
theories with only two physical DOFs in the unitary gauge.
(See Refs.~\cite{Afshordi:2014qaa,deRham:2016ged,Lin:2017oow,Chagoya:2018yna}
for related theories
developed with different motivations from
ours. We discuss the relation of their models to ours in \S \ref{ssec:comparison}.)
To this end,
we start from some scalar-tensor theory with three DOFs in general,
and identify the specific forms of the free functions in the Lagrangian by
requiring that the theory actually has only two DOFs.
Specifically, we proceed step by step
in the following way. First, we specify the Lagrangian having
the properties~[\ref{charA}] and [\ref{charB}].
This step can be done relatively easily, but
the two-DOF nature is ensured {\em only on a cosmological background}.
In this sense,
the properties~[\ref{charA}] and [\ref{charB}] are
just necessary conditions for the theory we aim to construct, and hence
the resultant Lagrangian should be considered as
a prototype. Next, we identify which of the theory among
this ``cosmological cuscuton'' class
has two DOFs {\em on an arbitrary background} in the unitary gauge.
If one starts from the k-essence theory and follows the above steps, one
arrives at the original cuscuton theory~\eqref{originalcuscuton}.
In this paper, we start from the GLPV theory and derive what we call the
{\em extended cuscuton} by this procedure.
We believe that the same procedure can in principle be applied to
even broader classes such as DHOST theories as a starting point, which we
hope to discuss in the near future.

The rest of this paper is organized as follows.
In \S \ref{sec:formulation}, we construct a prototype for
the extended cuscuton theory as a subclass of the GLPV theory
which has two DOFs at least on a cosmological background.
Then, in \S \ref{sec:Hamiltonian}, we perform a nonlinear Hamiltonian analysis
of the prototype models on an arbitrary background and identify the theory
with only two propagating DOFs, which yields our desired extended cuscuton theory.
The relation between the original and the extended cuscuton theories is studied by means of disformal transformation (i.e., a redefinition of the metric
which depends on $\phi$ and its first derivative~\cite{Bekenstein:1992pj}) in \S \ref{sec:disformaltransf}.
We also analyze cosmological perturbations in this theory in the presence of a matter field in \S \ref{sec:cospert}.
Finally, we draw our conclusions in \S \ref{sec:conc}.

\section{Cosmological prototype for extended cuscuton}
\label{sec:formulation}

The aim of the present paper is to find a general class of scalar-tensor theories with two physical DOFs in the unitary gauge.
As a first step to achieve this, in this section, we specify a subclass of
the GLPV theory with the aforementioned properties~[\ref{charA}] and [\ref{charB}] which characterize the cuscuton theory.

Our starting point is the GLPV theory~\cite{Gleyzes:2014dya}, whose action is given by
	\be
	S_{\rm GLPV}=\int d^4x\sqrt{-g}\bra{L_2^{\rm H}+L_3^{\rm H}+L_4^{\rm H}+L_5^{\rm H}+L_4^{\rm bH}+L_5^{\rm bH}}, \label{GLPVcov}
	\ee
where the first four terms form the Horndeski Lagrangian:
	\be
	\begin{split}
	L_2^{\rm H}&=G_2(\phi,X),\\
	L_3^{\rm H}&=G_3(\phi,X)\Box\phi,\\
	L_4^{\rm H}&=G_4(\phi,X)\mR+G_{4X}\brb{\bra{\Box\phi}^2-\phi_\mu^\nu\phi_\nu^\mu},\\
	L_5^{\rm H}&=G_5(\phi,X)\mG^{\mn}\phi_\mn-\fr{1}{6}G_{5X}\brb{\bra{\Box\phi}^3-3\bra{\Box\phi}\phi_\mu^\nu\phi_\nu^\mu +2\phi_\mu^\nu\phi_\nu^\la\phi_\la^\mu},
	\end{split}
	\ee
with $\mG_\mn$ being the Einstein tensor,
and the last two
are the beyond Horndeski terms:
	\be
	\begin{split}
	L_4^{\rm bH}&=F_4(\phi,X)\brc{-2X\brb{\bra{\Box\phi}^2-\phi_\mu^\nu\phi_\nu^\mu} -2\phi_\mu\phi^\mu_\nu\bra{\phi^\nu\Box\phi-\phi^\nu_\la\phi^\la}},\\
	L_5^{\rm bH}&=F_5(\phi,X)\,\Bigl\{-2X\brb{\bra{\Box\phi}^3-3\bra{\Box\phi}\phi_\mu^\nu\phi_\nu^\mu +2\phi_\mu^\nu\phi_\nu^\la\phi_\la^\mu} \\
	&~~~~~~~~~~~~~~~~~~-3\phi_\la\phi^\la_\si\phi^\si\brb{\bra{\Box\phi}^2-\phi_\mu^\nu\phi_\nu^\mu} +6\phi_\mu\phi^\mu_\nu\phi^\si\bra{\phi^\nu_\si\Box\phi-\phi^\nu_\la\phi^\la_\si}\Bigr\}.
	\end{split}
	\ee
Here, $G_2, G_3, G_4, G_5, F_4$, and $F_5$ are arbitrary functions of $(\phi,X)$, $\phi_{\mu}\equiv\na_{\mu}\phi$, and $\phi_\mn\equiv \na_\mu\na_\nu\phi$.

Now we consider a homogeneous and isotropic universe:
	\be
	ds^2=-N^2(t)dt^2+a^2(t)\delta_{ij}dx^idx^j,\quad \phi=\phi(t).
	\ee
The field equations are obtained
by substituting this ansatz into the action~\eqref{GLPVcov} and varying it with respect to $N$, $a$, and $\phi$.
Thereafter, we may set $N=1$.\footnote{Alternatively, even if one sets $N=1$ at the action level and then varies the action with respect to $a$ and $\phi$, the correct dynamical equations~\eqref{EOMa} and \eqref{EOMphi} are obtained. However, in this case, one cannot reproduce the Euler-Lagrange equation for $N$
from the dynamical equations~\cite{Motohashi:2016prk}.}
The structure of the two dynamical equations are as follows (see, e.g., Ref.~\cite{Kobayashi:2014ida}):
	\begin{align}
	\mE_a&=2\mG_T\dot{H}-2\mM\ddot{\phi}+\mU
  			=0, \label{EOMa} \\
	\mE_\phi&=6\mM\dot{H}+\mK\ddot{\phi}+\mV
  			=0, \label{EOMphi}
	\end{align}
where $H\equiv \dot a/a$
denotes the Hubble parameter and we have defined the following quantities:
	\be
	\begin{split}
	\mG_T&\equiv 2\bra{G_4-2XG_{4X}+XG_{5\phi}-H\dot{\phi}XG_{5X}+4X^2F_4-12H\dot{\phi}X^2F_5}, \\
	\mM&\equiv -XG_{3X}-G_{4\phi}-2XG_{4\phi X}+2H\dot{\phi}\bra{G_{4X}+2XG_{4XX}-G_{5\phi}-XG_{5\phi X}-8XF_4-4X^2F_{4X}} \\
	&~~~~~+H^2X\bra{3G_{5X}+2XG_{5XX}+60XF_5+24X^2F_{5X}}, \\
	\mK&\equiv G_{2X}+2XG_{2XX}+2\bra{G_{3\phi}+XG_{3\phi X}}-6H\dot{\phi}\bra{G_{3X}+XG_{3XX}+3G_{4\phi X}+2XG_{4\phi XX}} \\
	&~~~~~~+6H^2\,\bigl(G_{4X}+8XG_{4XX}+4X^2G_{4XXX}-G_{5\phi}-5XG_{5\phi X} \\
	&~~~~~~~~~~~~~~~~~~-2X^2G_{5\phi XX}-24XF_4-36X^2F_{4X}-8X^3F_{4XX}\bigr) \\
	&~~~~~~+2H^3\dot{\phi}\bra{3G_{5X}+7XG_{5XX}+2X^2G_{5XXX}+120XF_5+132X^2F_{5X}+24X^3F_{5XX}}, \\
	\mU&\equiv G_2+2XG_{3\phi}+4XG_{4\phi\phi}+4H\dot{\phi}\bra{G_{4\phi}-2XG_{4\phi X}+XG_{5\phi\phi}+4X^2F_{4\phi}} \\
	&~~~~~~+2H^2\bra{3G_4-6XG_{4X}+3XG_{5\phi}-2X^2G_{5\phi X}+12X^2F_4-24X^3F_{5\phi}} \\
	&~~~~~~-4H^3\dot{\phi}\,\bigl(XG_{5X}+12X^{2}F_{5}\bigr), \\
	\mV&\equiv -G_{2\phi}+2XG_{2\phi X}+2XG_{3\phi\phi}+3H\dot{\phi}\bra{G_{2X}+2G_{3\phi}-2XG_{3\phi X}-4XG_{4\phi\phi X}} \\
	&~~~~~~-6H^2\bra{3XG_{3X}+2G_{4\phi}+6XG_{4\phi X}-4X^2G_{4\phi XX}+XG_{5\phi\phi}-3H^{2}XG_{5X}-2H^{2}X^2G_{5XX} \right. \\ 
	&\left.~~~~~~~~~~~~~~~~~~+2X^2G_{5\phi\phi X}+12X^2F_{4\phi}+8X^3F_{4\phi X}-60H^{2}X^{2}F_{5}-24H^{2}X^{3}F_{5X}} \\
	&~~~~~~+2H^3\dot{\phi}\,\bigl(9G_{4X}+18XG_{4XX}-9G_{5\phi}-7XG_{5\phi X}+2X^2G_{5\phi XX} \\
	&~~~~~~~~~~~~~~~~~~-72XF_4-36X^2F_{4X}+48X^2F_{5\phi}+24X^3F_{5\phi X}\bigr).
\end{split} \label{def:GTetal}
	\ee
These quantities contain at most first derivatives of the scalar field and the metric.

In the case of the k-essence theory, we have $G_3=G_5=0$, $G_4=\,$const,
and hence ${\cal M}=0$. Then, the property~[\ref{charA}] reads
\begin{align}
{\cal K}&=G_{2X}+2XG_{2XX}=0
\notag \\ \Rightarrow\;\;\; G_2&=c_1(\phi)\sqrt{|X|}+c_2(\phi).
\end{align}
The original cuscuton theory~(\ref{originalcuscuton}) is thus recovered.
However, we have ${\cal M}\neq 0$ in general, which signals a kinetic mixing of gravity
and the scalar field. In this case, the statement of~[\ref{charA}] is
subtle, and instead it is more appropriate to require the following
extended version of~[\ref{charA}]:
\begin{enumerate}
\renewcommand{\theenumi}{\Alph{enumi}$'$}
\renewcommand{\labelenumi}{[\theenumi]}
\item \label{charAprime} \textit{The system composed of the two dynamical
equations~\eqref{EOMa} and \eqref{EOMphi} is degenerate}:
$\mG_T\mK+6\mM^2=0$.
\end{enumerate}
This condition can be rearranged to give
	\be
	\mG_T\mK+6\mM^2= \sum_{n=0}^4a_n(\phi,\dot{\phi})H^n=0,
	\ee
where $a_n$'s are functions of $\phi$ and $\dot{\phi}$.
The property~[\ref{charAprime}] is satisfied if
	\be
	a_n=0~~~(n=0,1,2,3,4), \label{cusc_cond_cov}
	\ee
which may be regarded as a set of differential equations
satisfied by $G_2$, $G_3, \cdots$ of the extended cuscuton.

Now let us move to the property~[\ref{charB}].
Following the standard procedure it is straightforward to derive the quadratic action
for the curvature perturbation $\zeta$ in the GLPV theory
(see, e.g., Ref.~\cite{Gleyzes:2014dya}). We have
\be
S_S^{(2)}=\int dtd^3xNa^3\brb{\mG_S\dot{\zeta}^2-\fr{\mF_S}{a^2}(\pa_k\zeta)^2}, \label{qac_s}
\ee
where it is found that
\begin{align}
{\cal G}_S\propto \mG_T\mK+6\mM^2.
\end{align}
Therefore, the two requirements~[\ref{charA}] and [\ref{charB}] are in fact equivalent.

Although Eq.~\eqref{cusc_cond_cov} provides some restrictions on
the functions in the GLPV action~\eqref{GLPVcov} and one
can specify the subclass satisfying this in principle, the actual manipulation is tedious.
To bypass this nonessential issue,
we move to the Arnowitt-Deser-Misner (ADM) formalism
rather than sticking to the covariant formulation.
It turns out that the ADM formalism greatly simplifies the analysis.

The GLPV action~\eqref{GLPVcov} is translated to the ADM language as follows:
    \begin{align}
    S_{\rm GLPV}=\int dtd^3xN\sqrt{\ga}\biggl[&A_2+A_3K+A_4(K^2-K^i_jK^j_i)+B_4R \nonumber \\
    &+A_5(K^3-3KK^i_jK^j_i+2K^i_jK^j_kK^k_i)+B_5
    \left(R^{ij}K_{ij}-\fr{R}{2}K\right)
    \biggr], \label{GLPVinADM}
    \end{align}
where we have taken the unitary gauge~$\phi=\phi(t)$.
Here,
$K_{ij}$ and $R_{ij}$ are the extrinsic and intrinsic curvature
tensors of $t=\,$const~hypersurfaces,
$K\equiv K^i_i$, $R\equiv R_i^i$,
and the coefficients~$A_2,A_3,A_4,A_5,B_4$, and $B_5$ are functions of $(t,N)$.
Indeed, in the unitary gauge, we have $X=\dot\phi^2(t)/(2N^2)$,
and hence a function of $(\phi,X)$ is mapped to
a function of $(t,N)$.
The relation between the two sets of the functions, $(G_i,F_j)$ in Eq.~\eqref{GLPVcov} and $(A_i,B_j)$ in Eq.~\eqref{GLPVinADM}, is given in Ref.~\cite{Gleyzes:2014dya}.
In the case of the Horndeski theory, only four of these six functions are independent,
as there exist the following constraints:
    \be
    A_4=-B_4-NB_{4N},\quad A_5=\fr{N}{6}B_{5N}, \label{Horndeski-tuning}
    \ee
where a subscript $N$ denotes $\pa/\pa N$.


In terms of $(A_i,B_j)$ instead of $(G_i,F_j)$, $a_n$ can be expressed as
\begin{align}
    a_{0}&\propto3\bra{A_{3}'}^{2} - 4\bra{A_{2}' + A_{2}''}A_{4},
    \label{eq:a0}\\
    a_{1}&\propto\bra{A_{3}'+ A_{3}''}A_{4} - 2A_{3}'A_{4}'  + \bra{A_{2}'+ A_{2}''}A_{5},
    \label{eq:a1}\\
    a_{2}&\propto2\bra{A_{4}'+ A_{4}''}A_{4} - 4\bra{A_{4}'}^{2} + 3\bra{A_{3}'+ A_{3}''}A_{5} - 3A_{3}'A_{5}',\label{eq:a2}\\
    a_{3}&\propto3\bra{A_{4}'+ A_{4}''}A_{5} - 6A_{4}'A_{5}' + A_{4}\bra{A_{5}' + A_{5}''},
    \label{eq:a3}\\
    a_{4}&\propto3\bra{A_{5}'}^{2} - 2\bra{A_{5}' + A_{5}''}A_{5},\label{eq:a4}
\end{align}
where $'\equiv\pa/\pa \ln N$.
In the following, we solve the system of differential equations $a_n=0$
to obtain the prototype of the extended cuscuton.
Since the structure of the system is different for $A_5=0$ and $A_5\neq 0$,
we treat these two cases separately.
It is worth noting that the coefficients $a_n$ are independent of $B_4$ and $B_5$.
This in particular means that no restrictions on $B_4$ and $B_5$
can be imposed from the analysis of the cosmological setup.

It should be noted finally that the condition $a_n=0$
is a sufficient but not a necessary condition for $\mG_S=0$: There is still a possibility that $\mG_S$ vanishes after imposing the Hamiltonian constraint for the background.
This is indeed the case in theories that are generated from the original cuscuton theory via generic disformal transformation (see \S \ref{sec:disformaltransf}).

\subsection{$A_5=0$ (and $A_4\ne 0$)}\label{ssec:A5=0}
In this case, $a_4=0$ and $a_3=0$ are automatically satisfied.
From $a_2=0$, we obtain
	\be
	A_4=-\fr{v_4N}{N+u_4},
	\ee
with $u_4$ and $v_4$ being arbitrary integration functions of $t$.
Hereafter, we assume $v_4\ne 0$ so that $A_4\ne 0$.
Then, $a_1=0$ yields
	\be
	A_3=u_3+\fr{v_3}{N+u_4} \label{A3(A5=0)},
	\ee
and $a_0=0$ can be solved to give
	\be
	A_2=u_2+\fr{v_2}{N}-\fr{3v_3^2}{8v_4N(N+u_4)},
	\ee
where $u_2,u_3,v_2$, and $v_3$ are arbitrary functions of $t$.
Since $u_3$ in Eq.~\eqref{A3(A5=0)} can be absorbed into $v_2$ through integration by
parts [see the form of the Lagrangian~\eqref{GLPVinADM}], we take $u_3=0$ from the beginning.
Thus, we have obtained
	\be
	A_5=0,\quad
  A_4=-\fr{v_4N}{N+u_4},\quad
  A_3=\fr{v_3}{N+u_4},\quad
  A_2=u_2+\fr{v_2}{N}-\fr{3v_3^2}{8v_4N(N+u_4)}. \label{ex_cusc(A5=0)}
	\ee

\subsection{$A_5\ne 0$}\label{ssec:A5!=0}
In this case, $a_4=0$ leads to the following solution for $A_5$:
	\be
	A_5=\fr{\pm N^2}{(\mu_5N+\nu_5)^2},
	\ee
with $\mu_5$ and $\nu_5$ being arbitrary functions of $t$ that do not vanish simultaneously.
Throughout this section, double signs are in the same order. 
One can then successively solve $a_3=0$, $a_2=0$, and $a_1=0$ to obtain
	\be
	\begin{split}
	A_4&=\fr{N(\mu_4N+\nu_4)}{(\mu_5N+\nu_5)^2}, \\
	A_3&=\mu_3+\fr{\nu_3}{\mu_5N+\nu_5}\pm\fr{2(\mu_4N+\nu_4)^2}{3(\mu_5N+\nu_5)^2},\\
	A_2&=\mu_2+\fr{\nu_2}{N}\pm\fr{\nu_3(\mu_4N+\nu_4)}{N(\mu_5N+\nu_5)}+\fr{2(\mu_4N+\nu_4)^3}{9N(\mu_5N+\nu_5)^2},
	\end{split}
	\ee
where $\mu_2,\mu_3,\mu_4,\nu_2,\nu_3$, and $\nu_4$ are arbitrary functions of $t$.
Finally, $\nu_3=0$ is imposed from $a_0=0$, so that we now have
	\be
	A_5=\fr{\pm N^2}{(\mu_5N+\nu_5)^2},
  \quad
  A_4=\fr{N(\mu_4N+\nu_4)}{(\mu_5N+\nu_5)^2},
  \quad
  A_3=\mu_3\pm\fr{2(\mu_4N+\nu_4)^2}{3(\mu_5N+\nu_5)^2},\quad
  A_2=\mu_2+\fr{\nu_2}{N}+\fr{2(\mu_4N+\nu_4)^3}{9N(\mu_5N+\nu_5)^2}. \label{ex_cusc}
	\ee
Here, $\mu_3$ can be absorbed into $\nu_2$, but we avoid doing so for later convenience.
Note that one can take a smooth limit~$\mu_5\to0$ or $\nu_5\to0$ in Eq.~\eqref{ex_cusc}.
It should also be noted that the result of the case with $A_5=0$
can be reproduced by choosing the integration functions as
	\be
	\begin{split}
	&\nu_5=u_4\mu_5,\quad
  \mu_4=-v_4\mu_5^2,\quad
  \nu_4=\mp\fr{3v_3}{4v_4}-u_4v_4\mu_5^2,\quad
  \mu_3=\mp\fr{2}{3}v_4^2\mu_5^2, \\ 
	&\mu_2=u_2+\fr{2}{9}v_4^3\mu_5^4,\quad
  	\nu_2=v_2\pm\fr{1}{2}v_3v_4\mu_5^2+\fr{2}{9}u_4v_4^3\mu_5^4,
	\end{split}
	\ee
and then taking the limit $\mu_5\to\infty$.

\section{Extended cuscuton from Hamiltonian analysis}
\label{sec:Hamiltonian}

Having constructed the cosmological
prototype of the extended cuscuton theory in the previous section,
now we perform its Hamiltonian analysis to identify
the theories truly
having two DOFs in the unitary gauge without any assumption on the underlying spacetime.

\subsection{General discussion}\label{ssec:generaldiscuss}

Before proceeding to the Hamiltonian analysis of the cosmological prototype of the
extended cuscuton,
we derive a (sufficient) condition for a theory written
in the ADM language to have DOFs less than three.
We start from a general ADM action of the form
	\be
	S=\int dtd^3x\,
  N\sqrt{\ga}\brb{L(t,N,\ga_{ij},R_{ij},Q_{ij})+v^{ij}(Q_{ij}-K_{ij})}, \label{general}
	\ee
respecting the three-dimensional spatial diffeomorphism invariance,
and explore the condition for $L$ to yield two DOFs.
Here, we have introduced Lagrange multipliers~$v^{ij}$ to replace $K_{ij}$ in $L$ by auxiliary variables~$Q_{ij}$.
This is thought of as the ADM expression of general scalar-tensor
theories in the unitary gauge.
Note that some DHOST theories yield the velocity of the lapse function~$\dot{N}$~\cite{Langlois:2017mxy}, which is beyond the scope of this paper.
We shall revisit the Hamiltonian structure when $L$ is at most quadratic in $K_{ij}$ in Appendix~\ref{app:A5=0}, which is the case for the extended cuscuton theory with $A_5=0$.

Switching to the Hamiltonian formalism, there are 44 canonical variables:
	\be
	\begin{pmatrix}
	N,&N^i,&\ga_{ij},&Q_{ij},&v^{ij}\\
	\pi_N,&\pi_i,&\pi^{ij},&P^{ij},&U_{ij}
	\end{pmatrix}.
	\ee
From the action~\eqref{general}, we obtain the primary constraints as
	\be
	\pi_N\approx0,\quad
  \pi_i\approx0,\quad
  P^{ij}\approx0,\quad
  U_{ij}\approx0,\quad
  \Psi^{ij}\equiv \pi^{ij}+\fr{\sqrt{\ga}}{2}v^{ij}\approx0.
	\label{primary}
	\ee
We will use the following notations for derivatives of $L$ with respect to $Q_{ij}$:
	\be
	L^{ij}_Q\equiv \fr{\pa L}{\pa Q_{ij}},\quad
  L^{ij,kl}_{QQ}\equiv \fr{\pa^2 L}{\pa Q_{ij}\pa Q_{kl}}.
	\ee
The canonical Hamiltonian can be obtained in the standard manner as
	\be
	H=\int d^3x\bra{N\mH_0+N^i\mH_i},
	\ee
with
	\be
	\mH_0\equiv -\sqrt{\ga}L(t,N,\ga_{ij},R_{ij},Q_{ij})+2\pi^{ij}Q_{ij},\quad
  \mH_i\equiv -2\sqrt{\ga}D^j\bra{\fr{\pi_{ij}}{\sqrt{\ga}}},
	\label{H0,Hi}
	\ee
where $D_i$ is the three-dimensional spatial covariant derivative.
The total Hamiltonian is written as
	\be
	H_T=H+\int d^3x(\la_N\pi_N+\la^i\pi_i+\chi_{ij}P^{ij}+\vp^{ij}U_{ij}+\la_{ij}\Psi^{ij}). \label{totalHamiltonian}
	\ee
Some of the consistency relations for the primary constraints produce
the following secondary constraints:
	\be
	\begin{split}
	\dot{\pi}_N&\approx\sqrt{\ga}(NL)_N-2\pi^{ij}Q_{ij}\equiv \mC\approx 0, \\
	\dot{\pi}_i&\approx -\mH_i\approx 0, \\
	\dot{P}^{ij}&\approx N(\sqrt{\ga}L^{ij}_Q-2\pi^{ij})\equiv N\Pi^{ij}\approx 0,
	\end{split}
	\ee
while $\dot U_{ij}\approx 0$ and $\dot \Psi^{ij}\approx 0$ just fix the multipliers
$\la_{ij}$ and $\vp^{ij}$, respectively.
The consistency relation from the time evolution of the secondary constraint $\mH_i\approx 0$, i.e., $\dot{\mH}_i\approx 0$, is automatically satisfied on the constraint surface.
Among the constraints derived so far, $\pi_i\approx 0$ is first class, which reflects the fact that one can freely specify the shift vector.
The momentum constraint~$\mH_i\approx 0$ can be promoted to a first-class constraint by adding appropriate terms that vanish weakly, i.e.,
	\be
	\mH_i~~~\to~~~\bar{\mH}_i\equiv \mH_i+\pi_ND_iN+P^{jk}D_iQ_{jk}-2\sqrt{\ga}D_j\bra{\fr{P^{jk}}{\sqrt{\ga}}Q_{ik}},
	\ee
so that $\bar{\mH}_i$ defines the generator of spatial diffeomorphisms for $\ga_{ij}$, $N$, and $Q_{ij}$.

Now we proceed to the consistency relations for $\mC\approx 0$ and $\Pi^{ij}\approx 0$.
One finds
	\begin{align}
	\dot{\mC}&\approx \pb{\mC,H}+\sqrt{\ga}\brb{\la_N (NL)_{NN}+\chi_{kl}NL^{kl}_{QN}}\approx 0, \label{3c1}\\
	\dot{\Pi}^{ij}&\approx \pb{\Pi^{ij},H}+\sqrt{\ga}\brb{\la_N L^{ij}_{QN}+\chi_{kl}L^{ij,kl}_{QQ}}\approx 0\label{3c2}.
	\end{align}
Therefore, if the matrix
	\be
	M\equiv \begin{pmatrix}
	(NL)_{NN}&NL^{kl}_{QN}\\
	L^{ij}_{QN}&L^{ij,kl}_{QQ}
	\end{pmatrix}
	\ee
has a nonvanishing determinant, the above consistency relations fix $\la_N$ and $\chi_{kl}$,
 and the Poisson algebra closes here.
If this is the case, we would
have 6 first-class and 26 second-class constraints, resulting in three DOFs.
Hence, we require
	\be
	\det M=\bra{\det L^{ij,kl}_{QQ}}\brb{(NL)_{NN}-NL^{ij}_{QN}(L^{-1}_{QQ})_{ij,kl}L^{kl}_{QN}}=0,
  \label{storngequaltiy}
	\ee
so that the theory~\eqref{general} has DOFs less than three.
Note that
this requirement might be too strong for the absence of the third DOF,
because it should be sufficient that $\det M$ vanishes only weakly,
$\det M\approx 0$ (see \S \ref{sec:disformaltransf}).
Nevertheless, in this paper, we require the
presumably stronger condition~\eqref{storngequaltiy} for simplicity.
Assuming $\det L^{ij,kl}_{QQ}\ne0$ to guarantee the existence of two propagating tensor DOFs, the above requirement reads
	\be
	\Delta\equiv (NL)_{NN}-NL^{ij}_{QN}(L^{-1}_{QQ})_{ij,kl}L^{kl}_{QN}=0. \label{2DOF-cond}
	\ee
Then, combining Eqs.~\eqref{3c1} and~\eqref{3c2} we obtain the tertiary constraint
	\be
	\Xi\equiv\pb{\mC,H}-N\pb{\Pi^{ij},H}(L^{-1}_{QQ})_{ij,kl}L^{kl}_{QN}\approx 0.
	\ee

Since the manipulations required hereafter are quite involved,
we only present a brief analysis.
The time evolution of the tertiary constraint will produce
the quaternary constraint: $\dot\Xi\approx0 \Rightarrow \Phi\approx 0$,
because otherwise
the number of phase-space dimensions would be odd and
the theory would be inconsistent.
Finally, the consistency relation~$\dot\Phi\approx 0$ will fix the multiplier~$\la_N$.
As we have two more second-class constraints than what we would have
in the $\Delta\ne 0$ case, the system has only two physical DOFs.\footnote{There may be another possibility for the system to have two physical DOFs: If $\Xi\approx 0$ is automatically satisfied by the existing primary/secondary constraints, then $\pi_N\approx 0$ and $\mC\approx 0$ should be first-class constraints and thus the number of DOFs is again two. In any case, $\Delta=0$
is a sufficient condition for the theory to have DOFs less than three.}

\subsection{The form of $A_i$ and $B_j$ satisfying the condition~\eqref{2DOF-cond}}\label{ssec:specificforms}

In the previous section, we have
obtained the candidate of the extended cuscuton theory
from the cosmological considerations.
In particular, recall that $B_4$ and $B_5$ are completely free
at this stage.
We now check whether or not
the candidate can satisfy the condition~\eqref{2DOF-cond}.
For theories whose action can be written in the form~\eqref{GLPVinADM}, we have
	\be
	\begin{split}
	(NL)_{NN}&=(NA_2)_{NN}+(NA_3)_{NN}Q+(NA_4)_{NN}\mQ_2+(NA_5)_{NN}\mQ_3+(NB_4)_{NN}R+(NB_5)_{NN}\left(R^{ij}Q_{ij}-\fr{R}{2}Q\right), \\
	L^{ij}_{QN}&=(A_{3N}+2A_{4N}Q+3A_{5N}\mQ_2)\ga^{ij}-(2A_{4N}+6A_{5N}Q)Q^{ij}+A_{5N}Q^i_kQ^{kj}+B_{5N}\left(R^{ij}-\fr{R}{2}\ga^{ij}\right), \\
	L^{ij,kl}_{QQ}&=-(2A_4+6A_5Q)\mG^{ij,kl}+6A_5\bra{Q^{k(i}\ga^{j)l}+Q^{l(i}\ga^{j)k}-Q^{ij}\ga^{kl}-\ga^{ij}Q^{kl}},
\end{split} \label{parts3}
	\ee
where $\mG^{ij,kl}\equiv \ga^{k(i}\ga^{j)l}-\ga^{ij}\ga^{kl}$ is the DeWitt metric and
	\be
	Q\equiv Q^i_i,\quad
  \mQ_2\equiv Q^2-Q^i_jQ^j_i,\quad
  \mQ_3\equiv Q^3-3QQ^i_jQ^j_i+2Q^i_jQ^j_kQ^k_i.
	\ee
The inverse of $L^{ij,kl}_{QQ}$ can be written as
\begin{align}
(L_{QQ}^{-1})_{ij,kl}=&-\frac{1}{2A_4}\left(
\ga_{k(i}\ga_{j)l}-\fr{1}{2}\ga_{ij}\ga_{kl}
\right) \nonumber \\
&+
\fr{3A_5}{4A_4^2}\brb{\bra{2\ga_{k(i}\ga_{j)l}-\ga_{ij}\ga_{kl}}Q+\ga_{ij}Q_{kl}+\ga_{kl}Q_{ij}-2\ga_{k(i}Q_{j)l}-2\ga_{l(i}Q_{j)k}}
+\cdots,
\label{generalinvert}
\end{align}
where the ellipsis denotes the terms quadratic and higher in $Q_{ij}$.
Thus, we obtain the equation of the form
\begin{align}
\Delta = \ti{c}_0(t,N)+\ti{c}_1(t,N)Q+\cdots +\ti{d}_1(t,N)R
+\ti{d}_2(t,N)\left(R_{ij}R^{ij}-\frac{3}{8}R^2\right)
+\ti{d}_3(t,N)QR+\cdots =0,
\end{align}
and all the coefficients must vanish. Here,
the $\ti{d}_i$ coefficients contain $B_4$ and $B_5$.
We see that $\ti{d}_2\propto (B_{5N})^2=0\;\Rightarrow\;B_5=b_2(t)$.
Then, $\ti{d}_1\propto (NB_4)_{NN}=0\;\Rightarrow\;B_4=b_0(t)+b_1(t)/N$.
However, $b_2$ can be absorbed into the redefinition of $b_1$.
We thus arrive at
	\be
	B_4=b_0(t)+\fr{b_1(t)}{N},\quad B_5=0, \label{condB}
	\ee
with $b_0$ and $b_1$ being free functions of $t$.
Now $B_4$ and $B_5$ are found to be eliminated from Eq.~\eqref{parts3} and $\Delta$,
and hence all the $\ti{d}_i$ coefficients vanish.

Let us then check that the form of $A_i$ we have found in the previous section
is consistent with the condition $\Delta=0$.


\subsubsection{$A_5=0$}\label{ssec:HA5=0}

Let us first take a look at the case with $A_5=0$,
for which simple explicit expressions of the equations can be obtained.
In this case, the inverse of the matrix $L^{ij,kl}_{QQ}=-2A_4\mG^{ij,kl}$
is given explicitly by
	\be
	(L^{-1}_{QQ})_{ij,kl}=-\frac{1}{2A_4}\left(
  \ga_{k(i}\ga_{j)l}-\fr{1}{2}\ga_{ij}\ga_{kl}
  \right),
	\ee
and hence we have
  \be
  \Delta=\frac{4(A_2'+A_2'')A_4-3(A_{3}')^2}{4NA_4}
  +\frac{(A_3'+A_3'')A_4-2A_{3}'A_{4}'}{NA_4}Q
  +\frac{(A_4'+A_4'')A_4-2(A_{4}')^2}{NA_4}\mQ_2, \label{det(A5=0)}
  \ee
  where recall that the prime denotes $\partial/\partial\ln N$.
As is clear from Eqs.~\eqref{eq:a0},~\eqref{eq:a1}, and~\eqref{eq:a2},
the three coefficients vanish if and only if $a_0=a_1=a_2=0$
(with $A_5=0$), and therefore $\Delta=0$
is satisfied for the functions~\eqref{ex_cusc(A5=0)}.

\subsubsection{$A_5\ne 0$}

In the $A_5\ne 0$ case,
one cannot express $L^{-1}_{QQ}$ in a closed form, but
rather one has an infinite sum of the form~\eqref{generalinvert}.
Then, we obtain $\Delta$ as
	\begin{align}
	\Delta=~&\fr{4(A_2'+A_2'')A_4-3(A_{3}')^2}{4NA_4}+\fr{4(A_3'+A_3'')A_4^2-8A_{3}'A_4A_{4}'+3A_3'^2A_{5}}{4NA_4^2}Q \nonumber \\
	&+\fr{8(A_4'+A_4'')A_4^3-16A_4^2(A_{4}')^2+12A_3'A_4(2A_4'A_5-A_4A_5')-9A_3'^2A_5^2}{8NA_4^3}\mQ_2 \nonumber \\
	&+\fr{8(A_5'+A_5'')A_4^4+3(3A_3'A_5-4A_4A_4')(4A_4^2A_5'-4A_4A_4'A_5+3A_3'A_5^2)}{8NA_4^4}\mQ_3 \nonumber \\
	&+\bra{2A_4^2A_5'-4A_4A_4'A_5+3A_3'A_5^2}^2\ti{\Delta}_{\ge 4}, \label{det(A5!=0)}
	\end{align}
where $\ti{\Delta}_{\ge 4}$ denotes higher-order terms of $Q_{ij}$.
It should be noted that this reduces to Eq.~\eqref{det(A5=0)} in the limit~$A_5\to0$.
Although Eq.~\eqref{det(A5!=0)} has infinitely many terms for generic choices of the
$A_i$ functions, one can check
directly that $\Delta = 0$ is satisfied if and only if
the $A_i$ functions are given by~\eqref{ex_cusc}.
\bigskip

Thus, we have established that the cosmological
prototype constructed in \S \ref{sec:formulation} can be promoted to
a theory with two DOFs, i.e., the extended cuscuton, by imposing the condition~\eqref{condB} on $B_4$ and $B_5$.
It turns out that we do not need to impose further constraints on the form of
the $A_i$ functions obtained from the cosmological analysis.
We present an alternative derivation of the same extended cuscuton in Appendix~\ref{app:nonflat}.
Given the action in the ADM form,
now it is straightforward to recast the theory to a covariant form via St\"{u}ckelberg trick, though the resultant expression is messy.
In Appendix~\ref{app:cov}, we present the expressions for $G_i(\phi,X)$ and $F_j(\phi,X)$ in
Eq.~\eqref{GLPVcov} for the extended cuscuton theory
with $A_5=0$.

In general, the extended cuscuton theory
contains a nonminimal derivative coupling to the curvature.
This is the reason why we have worked
in the GLPV framework.
The Horndeski conditions~\eqref{Horndeski-tuning}
are satisfied if and only if $A_5=0$, $u_4=0$, and $v_4=b_0(t)$.
Only in this case, the extended cuscuton theory can be
described as a special case of the Horndeski theory.

\subsection{Comparison with other related theories}\label{ssec:comparison}

We are now in a position to compare our extended cuscuton theory
with some other related theories in the literature.

The authors of Ref.~\cite{Afshordi:2009tt} extended the cuscuton theory to include
$G_3(\phi,X)\Box\phi$ to obtain consistently a generalization of the
McVittie solution. Their theory is included as a special case
in our extended cuscuton, but seemingly they have not addressed
the kinetic mixing of gravity and the scalar field
or the importance of the property~[\ref{charAprime}].
Another extension is the ``cuscuta-Galileon" proposed in Ref.~\cite{deRham:2016ged}.
This model is a subclass of the generalized Galileons in arbitrary dimensions that can avoid caustic singularities.
The cuscuta-Galileon is defined only in flat spacetime, so a direct comparison with our extended cuscuton would not be meaningful.
Yet another model was developed in Ref.~\cite{Chagoya:2018yna} as an extension of the Ho\v{r}ava-Lifshitz theory respecting the power-counting renormalizability.
This theory was shown to have two DOFs in the unitary gauge and it contains terms quadratic or higher in the curvature tensor, which are not incorporated in our extended cuscuton.
However, at the same time, there are many extended cuscuton models which do not fall into the theory studied in Ref.~\cite{Chagoya:2018yna}.

Besides the above concrete models, there are some general classes of two-DOF theories constructed in different ways than ours.
The authors of Ref.~\cite{Lin:2017oow} studied a class of theories depending on the lapse function at most linearly, i.e.,
	\be
	S=\int dtd^3x\,
	N\sqrt{\ga}L(t,\ga_{ij},R_{ij},K_{ij};D_i),
	\ee
and derived a condition on $L$ to yield two DOFs.
Although this theory generically lies outside our theory, it does not cover whole the extended cuscuton since our Lagrangian depends on $N$ nonlinearly.
In Ref.~\cite{Aoki:2018zcv}, another general class of scalar-tensor theories with two DOFs was invented by performing a canonical transformation on GR.
There should be some relation between this theory and ours, but the comparison would be far from trivial and thus we leave it for future work.

Finally, we would like to mention the relation between the work~\cite{DeFelice:2018ewo} and the present paper.
The authors of Ref.~\cite{DeFelice:2018ewo} studied a general class of scalar-tensor models where at most {\it three} DOFs propagate (i.e., no fourth DOF associated with Ostrogradsky instability) in the unitary gauge but the fourth DOF seemingly revives in other gauges, which was called ``U-degenerate'' theory.
They claimed that the fourth DOF actually does not propagate once a physically reasonable boundary condition at spatial infinity is imposed, and thus the U-degenerate theory is free of Ostrogradsky ghost as long as one can take the unitary gauge.
Now we see the similarity to our extended cuscuton:
The extended cuscuton theory exhibits the two-DOF nature
at least in the unitary gauge, but the situation may change
if one considers other gauges or the case where the unitary gauge
cannot be taken anyway.
As is the case for the U-degenerate theory, an appropriate
boundary condition at spatial infinity would kill the extra DOF in a generic gauge.
We address this issue in Appendix~\ref{app:nonU}.

\section{Disformal transformations}\label{sec:disformaltransf}
The original cuscuton model~\eqref{originalcuscuton} can be represented in the language of the GLPV action~\eqref{GLPVcov} as
	\be
	A_2=-V(\phi(t))+\fr{\sigma(t)}{N},\quad
  A_4=-B_4=-\fr{1}{2\kappa^2},\quad
  A_3=A_5=B_5=0, \label{ori_cusc}
	\ee
with $\sigma(t)\equiv \mu^{2}|\dot{\phi}(t)|$.
In this section, we study the behavior of the extended cuscuton theory under disformal transformation~\cite{Bekenstein:1992pj} and show that
a particular subclass with $A_5=0$ can be generated from the original cuscuton theory.

Let us consider (invertible) disformal transformation
$g_\mn \to\Om(t)g_\mn+\Ga(t,N)\phi_\mu\phi_\nu$ of
the original cuscuton model, with
	\be
	\Om=2\kappa^2 v_4,\quad
  \Ga=-\fr{\Om u_4}{\dot{\phi}^2}\bra{2N+u_4}. \label{disformaltrans}
	\ee
The above transformation contains two arbitrary functions, $u_4$ and $v_4$, of $t$.
Then, the original theory with the coefficients~\eqref{ori_cusc} is mapped to another GLPV theory with the following coefficients:
	\begin{align}
	&A_5=0,\quad
  A_4=-\fr{v_4N}{N+u_4},\quad
  A_3=\fr{v_3}{N+u_4},\quad
  A_2=u_2+\fr{v_2}{N}-\fr{3v_3^2}{8v_4N(N+u_4)}, \label{Adisf(A5=0)}\\
	&B_5=0,\quad
  B_4=v_4\bra{1+\fr{u_4}{N}}, \label{Bdisf(A5=0)}
	\end{align}
where $v_3$, $u_2$, and $v_2$ are given
by
	\be
	v_3=-2\dot{v}_4,\quad
  u_2=-\Omega^2V,\quad
  v_2=\Omega^{3/2}\sigma-\Omega^2u_4V.\label{condiuvu}
	\ee
These $A_i$ and $B_j$ functions are of the form of~\eqref{ex_cusc(A5=0)}
and~\eqref{condB}, but the $t$-dependent functions are subject to~\eqref{condiuvu}.
Therefore, the theory generated from the original cuscuton
via the disformal transformation~\eqref{disformaltrans}
resides in a particular subclass of the extended cuscuton theory.
The generated theory has two DOFs on any spacetime which is compatible with the unitary gauge.
This result is reasonable as an invertible disformal transformation does not change the number of physical DOFs~\cite{Domenech:2015tca,Takahashi:2017zgr}.

One could perform more general disformal transformations, but then the resultant theories generically lie beyond the current framework in the sense that
the condition~$\Delta=0$ for the absence of the third DOF (see \S \ref{sec:Hamiltonian}) is satisfied only {\it weakly}.
Although it may offer a possible generalization of the present formulation of
cuscuton theories retaining two DOFs,
we leave it for future study.


\section{Stability in the presence of matter}\label{sec:cospert}
In this section, we discuss the stability of cosmological solutions
in the extended cuscuton theory in the presence of a matter field, generalizing the result of~\cite{Boruah:2017tvg}.
We add a scalar field $\chi$ minimally coupled to gravity,
whose Lagrangian has the form
\begin{align}
	\mL_\chi=P(Y),\quad Y\equiv-\fr{1}{2}g^{\mu\nu}\pa_{\mu}\chi\pa_{\nu}\chi.
\end{align}
For simplicity, we assume that $P$ is a function of $Y$ and
does not depend on $\chi$ explicitly.
Such a scalar field can mimic a barotropic perfect fluid.
The energy density, pressure,  and sound speed of $\chi$ are respectively written as
\begin{align}
	\rho=2YP_{Y}-P,\quad p=P,\quad
  c_{s}^{2}=\fr{dp}{d\rho}=\fr{P_{Y}}{P_{Y} + 2YP_{YY}}.\label{matter}
\end{align}
Now we consider scalar perturbations around a cosmological background.
We choose the unitary gauge for the cuscuton field, $\phi=\phi(t)$, and write each constituent of the metric as
	\be
	N=1+\del N,\quad
  N_{i}=\pa_{i}\psi,\quad
  \ga_{ij}=a^{2}e^{2\zeta}\bra{e^h}_{ij}=a^{2}e^{2\zeta}\bra{\del_{ij} + h_{ij} + \fr{1}{2}h_{ik}h_{kj}+\cdots},
	\ee
where $\del N$, $\psi$, and $\zeta$ are scalar perturbations and $h_{ij}$ denotes transverse-traceless tensor perturbations.
The matter scalar field also fluctuates as $\chi= \chi(t) + \del \chi(t,\vec{x})$.

The quadratic action for the tensor perturbations is independent of the matter sector, which takes the form
	\be
	S_T^{(2)}=\fr{1}{8}\int dtd^3xa^3\brb{\mG_T\dot{h}_{ij}^2-\fr{\mF_T}{a^2}(\pa_k h_{ij})^2}, \label{qac_t}
	\ee
where
	\be
	\mG_T\equiv -2(A_4+3HA_5),\quad \mF_T\equiv 2B_4+\dot{B}_5. \label{coeff_t}
	\ee
Thus, the tensor perturbation $h_{ij}$ is stable if $\mG_T>0$ and $\mF_T>0$.
The equations are completely the same as in the GLPV theory
and we do not see any cuscuton nature at this point.

The quadratic Lagrangian for the scalar perturbations is $L^{(2)}=a^3\bra{\mL^{(2)}_H+\mL^{(2)}_\chi}$ with
	\begin{align}
	\mL^{(2)}_H&=-3\mG_T\dot{\zeta}^2+\fr{\mF_T}{a^2}\bra{\pa_k\zeta}^2+\Sigma\delta N^2
	-2\Theta\delta N\fr{\pa^2\psi}{a^2}+2\mG_T\dot{\zeta}\fr{\pa^2\psi}{a^2}+6\Theta\delta N\dot{\zeta}
	-2\bar{\mG}_T\delta N\fr{\pa^2\zeta}{a^2}, \\
	\mL^{(2)}_\chi&=\fr{P_Y}{c_s^2}\brb{-\fr{c_s^2}{2a^2}\bra{\pa_k\delta\chi}^2+c_s^2\dot{\chi}\fr{\pa^2\psi}{a^2}\delta\chi
	+Y\delta N^2-\dot{\chi}\bra{\delta N-3c_s^2\zeta}\dot{\delta\chi}+\fr{1}{2}\dot{\delta\chi}^2},
	\end{align}
where
	\be
	\begin{split}
	{\bar{\mG}_T}&\equiv 2(B_4+B_{4N})-HB_{5N}, \\
	\Sigma&\equiv A_{2N}+\fr{1}{2}A_{2NN}+\fr{3}{2}HA_{3NN}+3H^2\bra{2A_4-2A_{4N}+A_{4NN}}+3H^3\bra{6A_5-4A_{5N}+A_{5NN}}, \\
	\Theta&\equiv \fr{1}{2}A_{3N}-2H\bra{A_4-A_{4N}}-3H^2\bra{2A_5-A_{5N}}.
	\end{split}
	\ee
Note that $\mG_S$ in Eq.~\eqref{qac_s} can be written as $\mG_S=(\mG_T/\Theta^2)(\Sigma\mG_T+3\Theta^2)$, so the condition~$\mG_S=0$, which any cuscuton theory must satisfy (see \S \ref{sec:formulation}), implies
	\be
	\Sigma\mG_T+3\Theta^2=0. \label{Gs=0}
	\ee
Variations of $L^{(2)}$ with respect to the auxiliary variables~$\del N$ and $\psi$ yield
\begin{align}
\label{variationdelN}
	&\bra{\Sigma + \fr{YP_{Y}}{c_{s}^{2}}}\del N - \Theta\fr{\pa^{2}\psi}{a^{2}} + 3\Theta\dot{\zeta} - \mG_{T}\fr{\pa^{2}\zeta}{a^{2}}
		-\fr{\dot{\chi}P_{Y}}{c_{s}^{2}}\dot{\del\chi}=0,\\
	&\Theta\del N - \mG_{T}\dot{\zeta} - \frac{1}{2}\dot{\chi}P_{Y}\del \chi=0,
\end{align}
by which we can eliminate $\delta N$ and $\psi$ from $L^{(2)}$:
	\begin{align}
	L^{(2)}=a^3
  \left[\sizecorr{\bra{
  \dot{\zeta}
  -\fr{\Theta}{\mG_T}\fr{\dot{\delta\chi}}{\dot{\chi}}}^2}\right.
	&\fr{\mG_T^2YP_Y}{c_s^2\Theta^2}\bra{\dot{\zeta}-\fr{\Theta}{\mG_T}\fr{\dot{\delta\chi}}{\dot{\chi}}}^2
	-\fr{2(YP_Y)^2}{c_s^2\Theta}\fr{\dot{\delta\chi}\delta\chi}{\dot{\chi}^2}+\bra{\Sigma+\fr{YP_Y}{c_s^2}}\fr{YP_Y}{\Theta^2}
	\bra{2\mG_T\dot{\zeta}\fr{\delta\chi}{\dot{\chi}}+YP_Y\fr{\delta\chi^2}{\dot{\chi}^2}} \nonumber \\
	&-\fr{\mF_S}{a^2}\bra{\pa_k\zeta}^2+2{\bar{\mG}_T}\fr{YP_Y}{\Theta}\fr{\pa_k\zeta\pa_k\delta\chi}{a^2\dot{\chi}}
	-\fr{YP_Y}{a^2}\fr{(\pa_k\delta\chi)^2}{\dot{\chi}^2}
	\left.\sizecorr{\bra{\dot{\zeta}^2-\fr{\Theta}{\mG_T}\fr{\dot{\delta\chi}}{\dot{\chi}}}^2}\right],
	\end{align}
where we have defined
	\be
	\mF_S\equiv \fr{1}{a}\fr{d}{dt}\bra{\fr{a}{\Theta}\mG_T{\bar{\mG}_T}}-\mF_T,
	\ee
and used the background EOM for $\chi$, $\ddot{\chi}+3c_s^2H\dot{\chi}=0$.
One can remove the kinetic term for $\delta\chi$
by making the field redefinition
	\be
	\ti{\zeta}\equiv \zeta-\fr{\Theta}{\mG_T}\fr{\delta\chi}{\dot{\chi}}.
	\ee
Then,
$\delta\chi$ becomes an auxiliary variable and thus can be eliminated by using its EOM.
After tedious but straightforward manipulations, we finally arrive at
	\be
	L^{(2)}=a^3\brb{\mA(t,\partial^2)\dot{\ti{\zeta}}^2-\mB(t,\partial^2)
  \fr{(\pa_k\ti{\zeta})^2}{a^2}}, \label{LagF}
	\ee
where $\mA$ and $\mB$ are given respectively by
	\be
	\mA=\fr{\mG_T^2YP_Y}{c_s^2\Theta^2}\fr{\partial^2/a^2-\al_1}{\partial^2/a^2-\al_2},
  \quad
	\mB={\Upsilon}\fr{\mG_T^2YP_Y}{\Theta^2}
  \fr{\partial^4/a^4-\beta_1\partial^2/a^2+\beta_2}{(\partial^2/a^2-\al_2)^2}. \label{LagF_coeff}
	\ee
Here, we have defined
	\be
	\begin{split}
	\alpha_1&\equiv\fr{3}{f}\alpha_2,~~~\alpha_2\equiv-\fr{\bar{\Upsilon}^2c_s^2\Theta^2YP_Y}{\mF_S\Theta^2-\Upsilon \mG_T^2YP_Y}f(f-3), \\
	\beta_1&\equiv\alpha_2\bra{1+\fr{\mF_S\Theta^2}{\Upsilon\mG_T^2YP_Y}}-\fr{\Theta^2}{a^3\Upsilon\mG_T^2YP_Y}
	\fr{d}{dt}\brb{\fr{a^3\bar{\Upsilon}\mG_TYP_Y}{\Theta}(f-3)}, \\
	\beta_2&\equiv\fr{\Theta^2\alpha_2^2}{\Upsilon\mG_T^2YP_Y}\brc{\mF_S-\fr{1}{a}\fr{d}{dt}\brb{\fr{a\bar{\Upsilon}\mG_TYP_Y}{\Theta\alpha_2}(f-3)}},
	\end{split}
	\ee
with
	\be
	\begin{split}
	\Upsilon&\equiv \fr{\mF_S\Theta^2-\bar{\mG}_T^2YP_Y}{\mF_S\Theta^2-\mG_T(2\bar{\mG}_T-\mG_T)YP_Y}, \\
	\bar{\Upsilon}&\equiv \fr{\mF_S\Theta^2-\mG_T\bar{\mG}_TYP_Y}{\mF_S\Theta^2-\mG_T(2\bar{\mG}_T-\mG_T)YP_Y}, \\
	f&\equiv \fr{\mG_TYP_Y}{c_s^2\Theta^2}-\fr{1}{c_s^2}\fr{d}{dt}\bra{\fr{\mG_T}{\Theta}}+\fr{3\mG_TH}{\Theta}.
	\end{split}
	\ee
Thus, we have a single scalar DOF associated with
  the matter field. Interestingly,
the quadratic action is of a nonlocal form and as a result
  the dispersion relation is nonstandard.
  This means that the nature of scalar cosmological perturbations
  is different from that in GR in the presence of
  a perfect fluid. In other words, gravity is indeed modified in the cuscuton theory.
Note in passing that under the Horndeski tuning~\eqref{Horndeski-tuning},
$\mG_T$ and $\bar{\mG}_T$ coincide, and hence $\Upsilon=\bar{\Upsilon}=1$.

It follows that as long as
	\be
	\rho+p=2YP_Y>0,\quad c_s^2>0,\quad \Upsilon>0, \label{condforscalar}
	\ee
are satisfied, scalar perturbations are stable in the ultraviolet regime.
In the infrared regime, both ghost/gradient instabilities are not necessarily problematic:~Even if the kinetic term has a wrong sign, it is legitimate to ignore the ghost instability if its energy scale is much lower than the cutoff scale.
The gradient instability is also irrelevant when the timescale of interest is much shorter than that of the instability.
Note that the first two conditions are related only to the matter field,
stating that $\chi$ must be ``usual'' matter in the sense that it satisfies
the null energy condition and has a positive sound speed squared.
However, the last condition, $\Upsilon>0$, depends on the concrete form of the
cuscuton Lagrangian as well as the matter field, and hence is nontrivial.

\section{Summary and Discussion}\label{sec:conc}

The cuscuton theory is a special case of single-field scalar-tensor theories having only two DOFs, i.e., no propagating scalar DOF, in the unitary gauge.
Focusing on a cosmological setup,
the cuscuton exhibits the following properties:
[\ref{charA}] the field equation of the scalar field is at most of first order and
[\ref{charB}] the kinetic term of scalar cosmological perturbations vanishes.
In the present paper,
we have explored a possible extension of the cuscuton theory
in the context of the GLPV theory.
In doing so, the property [\ref{charA}] has been appropriately generalized
to the case with a kinetic mixing of gravity and the scalar field.
More specifically, [\ref{charAprime}] the system of the two dynamical equations
governing the background cosmological evolution is degenerate.

In \S \ref{sec:formulation}, we constructed the
cosmological prototype
of the extended cuscuton theory by imposing the conditions~[\ref{charAprime}]
and [\ref{charB}] on the GLPV action,
which are characterized by six free functions: $G_2$, $G_3$, $G_4$, $G_5$, $F_4$, and $F_5$
of $\phi$ and $X=-g^{\mu\nu}\phi_\mu\phi_\nu/2$
in the covariant form, or
$A_2$, $A_3$, $A_4$, $A_5$, $B_4$, and $B_5$ of
$t$ and the lapse function $N$
in the ADM representation.
It turned out that the conditions [\ref{charAprime}]
and [\ref{charB}] are in fact equivalent.
At this stage,
the $B_j$ functions remain arbitrary, while the $A_i$ functions are
fixed to be~\eqref{ex_cusc(A5=0)} in the $A_5=0$ case
and~\eqref{ex_cusc} in the $A_5\ne 0$ case.
Thereafter, to obtain the complete form of the
extended cuscuton theory, i.e., the theory having two physical DOFs on
any background spacetime under the unitary gauge, we performed a Hamiltonian analysis of the precursory models in \S \ref{sec:Hamiltonian}.
The requirement of having two DOFs poses a constraint~\eqref{condB} on the $B_j$
functions, and thus we obtained the desired extended cuscuton Lagrangian.


Furthermore, in \S \ref{sec:disformaltransf}, we studied the relation between the original and extended cuscuton theories by use of disformal transformation.
We showed that the theory that are mapped from the original cuscuton model by the disformal transformation~\eqref{disformaltrans} belong to
the $A_5=0$ case of our extended cuscuton theory.


We also studied scalar and tensor cosmological
perturbations in the presence of another scalar field
as matter in \S \ref{sec:cospert}.
The stability conditions for the tensor modes
are given by $\mG_T>0$ and $\mF_T>0$, where $\mG_T$ and $\mF_T$ are defined in Eq.~\eqref{coeff_t}.
These remain the same as the corresponding conditions in the GLPV theory.
The scalar modes acquire nonlocal interaction as in Eq.~\eqref{LagF} and
the stability conditions read Eq.~\eqref{condforscalar}.

Having formulated the extended cuscuton theory, it would be intriguing to study its phenomenological aspects such as early and late-time cosmology.
Black hole solutions in the extended cuscuton theory would be also interesting
to explore.
In parallel to phenomenology, we expect that
further extension of the cuscuton framework is still possible.
UV completion of the extended cuscuton theory is also an open question.
These issues will be addressed in forthcoming publications.


Before closing this final section, let us
comment on the constraint on the gravitational wave speed~$c_{\rm GW}$.
From the almost simultaneous detection of the gravitational waves GW170817 and the $\gamma$-ray burst GRB170817A~\cite{TheLIGOScientific:2017qsa,GBM:2017lvd,Monitor:2017mdv,Sakstein:2017xjx} from a binary neutron star merger, the deviation of $c_{\rm GW}$ from the speed of light
($c_{\rm light}\equiv 1$) is strongly constrained: $|c_{\rm GW}-1|\lesssim 10^{-15}$.
If one uses the extended cuscuton theory to modify gravity in the
present universe, this constraint must be respected. Therefore, here we present
the subclass of the extended cuscuton satisfying $c_{\rm GW}=1$ exactly.
In the GLPV theory satisfying this condition irrespective of the background spacetime,
the functions~$G_4$, $G_5$, $F_4$, and $F_5$ in Eq.~\eqref{GLPVcov} must obey~\cite{Ezquiaga:2017ekz,Creminelli:2017sry,Langlois:2017dyl}
	\be
	F_4=-8\fr{G_{4X}}{X},\quad G_5=F_5=0.
	\ee
Let us apply this requirement to the extended cuscuton Lagrangian.
Since $G_5=F_5=0$ implies $A_5=B_5=0$, we employ the case presented in \S \ref{ssec:A5=0} with Eq.~\eqref{condB}.
Then, imposing the condition $F_4=-8G_{4X}/X$ we obtain
	\be
	A_2=u_2+\fr{v_2}{N}-\fr{3v_3^2}{8v_4N^2},\quad A_3=\fr{v_3}{N},\quad A_4=-B_4=-v_4,
	\ee
in the ADM representation,
which is translated to the covariant form~\eqref{GLPVcov} with
	\be
	\begin{split}
	G_2&=\ti{u}_2+\ti{v}_2\sqx-\bra{2\ti{v}_3'+4\ti{v}_4''+\fr{3\ti{v}_3^2}{4\ti{v}_4}}X+\bra{\ti{v}_3'+2\ti{v}_4''}X\log X, \\
	G_3&=-\bra{\fr{\ti{v}_3}{2}+\ti{v}_4'}\log X,\quad G_4=\ti{v}_4,\quad G_5=F_4=F_5=0,
	\end{split}
	\ee
where $\ti{u}_2$, $\ti{v}_2$, $\ti{v}_3$, and $\ti{v}_4$ are arbitrary functions of $\phi$.
This is the same theory as the one studied in Ref.~\cite{Afshordi:2014qaa}.
All the other interactions introduced in our extended cuscuton theory
are strongly constrained by GW170817.
However, we emphasize that
this constraint applies only to the low-redshift universe ($z\lesssim 0.01$),
and a fairly large deviation of $c_{\rm GW}$ from unity may be possible
in the early universe.
Moreover, as has been pointed out recently in Ref.~\cite{deRham:2018red},
the energy scale which can be observed by LIGO
lies close to the cutoff scale of many dark energy models.
Hence, our extended cuscuton framework is worth investigating as
a model of cosmology and modified gravity.


\acknowledgements{
We would like to thank Katsuki Aoki and David Langlois for fruitful discussions.
This work was supported in part by
the Rikkyo University Special Fund for Research (A.I.),
JSPS Research Fellowships for Young Scientists No.~17J06778 (K.T.),
MEXT KAKENHI Grant Nos.~JP15H05888, JP16H01102, and JP17H06359 (T.K.),
JSPS KAKENHI Grant No.~JP16K17707 (T.K), and
MEXT-Supported Program for the Strategic Research Foundation at Private Universities, 2014-2018 (S1411024) (T.K.).
}


\appendix
\section{More on the Hamiltonian analysis in the $A_5=0$ case}\label{app:A5=0}

In this appendix, we examine the Hamiltonian structure of the extended cuscuton theory with
$A_5=0$ in more detail.
We will show that
(i) the Hamiltonian can be recast into the form in which
 the lapse function appears only linearly, as in the theories
 studied in Ref.~\cite{Lin:2017oow}, via canonical transformation;
 and that (ii) the analysis is rather simplified
 if we do not introduce the auxiliary variables~$Q_{ij}$ from the beginning.

For the extended cuscuton model with $A_5=0$,
the explicit form of the Lagrangian is given by
\begin{align}
	L=u_2 + \fr{v_2}{N} -\frac{3v_3^{2}}{8v_4N(N+u_4)} + \frac{v_3}{N + u_4}Q
		+\frac{v_4 N}{N + u_4}(Q^i_jQ^j_i-Q^2) + \bra{b_0+\fr{b_1}{N}}R,
\end{align}
plus the Lagrange multiplier term enforcing $Q_{ij}=K_{ij}$.
The subsequent analysis can be done
in the same way as in \S \ref{sec:Hamiltonian}.
Using the notation, the total Hamiltonian is given by
	\be
	\begin{split}
	H_T&=H+\int d^3x(\la_N\pi_N+\la^i\pi_i+\chi_{ij}P^{ij}+\vp^{ij}U_{ij}+\la_{ij}\Psi^{ij}), \\
	H&=\int d^3x \bra{-N\sqrt{\ga}L+2N\pi^{ij}Q_{ij}+2\pi^{ij}D_iN_j}.
	\end{split}
	\ee
This Hamiltonian depends nontrivially on $N$.
Now we perform the following canonical transformation:
	\be
	Q_{ij}\to \frac{N+u_{4}}{N}Q_{ij} + \frac{v_{3}}{4v_{4}N}\ga_{ij},\quad
	\ga_{ij}\to \ga_{ij},\quad
	P^{ij}\to \frac{N}{N+u_4}P^{ij},\quad
	\pi^{ij}\to \pi^{ij} - \frac{v_3}{4v_4(N+u_4)}P^{ij}.
	\ee
Then,
$H$ is transformed to
	\begin{align}
	H\to \int d^3x\,\biggl\{ &-\sqrt{\ga}\brb{Nu_2+v_2+(N+u_4)(Q^i_jQ^j_i-Q^2)+(Nb_0+b_1)R} \nonumber \\
	&+2\pi^{ij}\brb{(N+u_4)Q_{ij}+\fr{v_3}{4v_4}\ga_{ij}}+2\pi^{ij}D_iN_j\biggr\},
	\end{align}
  where
  the terms proportional to $P_{ij}$ were absorbed into the redefinition of $\chi_{ij}$.
Now we see that the new Hamiltonian
depends on $N$ at most linearly.

The analysis becomes simpler
if one does not employ auxiliary fields $Q_{ij}$ from the beginning.
Indeed, after straightforward calculations, the total Hamiltonian is obtained as
	\be
	\begin{split}
	H_T&=H+\int d^3x(\la_N\pi_N+\la^i\pi_i), \\
	H&=\int d^3x\,\brc{\sqrt{\ga}\brb{-Nu_2-v_2-(Nb_0+b_1)R+\fr{N+u_4}{2v_4}\fr{2\pi^i_j\pi^j_i-\pi^2}{\ga}+\fr{v_3}{2v_4}\fr{\pi}{\sqrt{\ga}}}+2\pi^{ij}D_iN_j},
	\end{split}
	\ee
where $\pi\equiv \pi^i_i$, and thus
it is found without invoking the canonical transformation
that the dependence of $H$ on $N$ is at most linear.

\section{Extended cuscuton from non-flat cosmology}
\label{app:nonflat}

In \S~\ref{sec:formulation}, we consider a cosmological background
to construct a prototype for the extended cuscuton.
By this approach, one cannot determine the form of $B_4$ and $B_5$
and their form was explored based on the Hamiltonian analysis in \S~\ref{sec:Hamiltonian}.
However, in this appendix, we show that
by considering a non-flat cosmological background
one can fix the form of $B_4$ and $B_5$ and the result agrees
with that obtained from the Hamiltonian analysis.

For a non-flat cosmological background with
\begin{align}
ds^2=-N^2dt^2+a^2\left[
\frac{dr^2}{1-kr^2}+r^2\left(d\theta^2+\sin^2\theta d\varphi^2\right)
\right],
\end{align}
the two dynamical equations take the same form as Eqs.~\eqref{EOMa} and~\eqref{EOMphi},
\begin{align}
{\cal E}_a&=2{\cal G}_T\dot H-2{\cal M}\ddot\phi+\cdots=0,
\\
{\cal E}_\phi&=6{\cal M}\dot H+{\cal K}\ddot\phi+\cdots=0,
\end{align}
but now with
\be
\begin{split}
{\cal G}_T&={\cal G}_{T\,{\rm flat}},
\\
{\cal M}&={\cal M}_{\rm flat}+XG_{5X}\frac{k}{a^2},
\\
{\cal K}&={\cal K}_{\rm flat}+6
\left[
G_{4X}+2XG_{4X}-G_{5\phi}-XG_{5\phi X}+H\dot\phi(G_{5X}+XG_{5XX})
\right]\frac{k}{a^2},
\end{split}
\ee
where the quantities labeled by ``flat'' represent
the corresponding ones in the flat case found in Eq.~\eqref{def:GTetal}.
This leads to
\begin{align}
{\cal G}_T{\cal K}+6{\cal M}^2=
\sum_{n=0}^4a_nH^n+
a_5\frac{k}{a^2}+a_6\frac{k^2}{a^4}
+a_7H\frac{k}{a^2}+a_8H^2\frac{k}{a^2},
\end{align}
where the coefficients of the four additional terms must vanish.

Switching from the $(G_i,F_j)$ representation to the $(A_i,B_j)$ representation,
first we see that
\begin{align}
a_6=6(XG_{5X})^2\propto (B_{5N})^2 = 0.
\end{align}
Substituting this to the other coefficients, we obtain
\begin{align}
a_5\propto A_4(NB_4)_{NN},
\quad
a_7\propto A_5(NB_{4})_{NN},
\quad
a_8=0.
\end{align}
We thus arrive at the same result as in the Hamiltonian analysis:
\begin{align}
(NB_{4})_{NN}=0,\quad B_{5N}=0.
\end{align}

\section{Covariantized form of the extended cuscuton}\label{app:cov}

In this appendix, we present the covariantized form of the extended cuscuton model with $A_5=0$.

To restore general covariance, we introduce a St\"{u}ckelberg field~$\phi$ so that its gradient is proportional to the unit normal vector to a constant-time hypersurface:~$n_\mu=-\phi_\mu/\sqx$~\cite{Blas:2009yd,Blas:2010hb}.
Then, the ingredients of the ADM action can be
rewritten in the following way:
	\be
	\begin{split}
	N&\to\fr{1}{\sqx},\quad
  \ga_{ij}\to h_\mn\equiv g_\mn+\fr{1}{2X}\phi_\mu\phi_\nu,
  \quad K_{ij}\to \mK_\mn\equiv h_\mu^\la\na_\la n_\nu, \\
	R_{ij}&\to h_\mu^\al h_\nu^\ga h^{\beta\delta}\mR_{\al\beta\ga\delta}-\mK^\al_\al\mK_\mn+\mK_\mu^\al\mK_{\al\nu},
	\end{split}
	\ee
while the functions of $t$ are replaced with
those of $\phi$:~$u_i(t)\to\ti{u}_i(\phi)$, $v_i(t)\to\ti{v}_i(\phi)$, and $b_i(t)\to\ti{b}_i(\phi)$.
The result is given by
	\be
	\begin{split}
	G_2&=\ti{u}_2+\ti{v}_2\sqx-4\ti{b}_0''X+2\ti{b}_1''(2X)^{3/2}
	-\fr{\ti{v}_3X}{1+\ti{u}_4\sqx}\bra{\fr{3\ti{v}_3}{4\ti{v}_4}+2\ti{u}_4'\sqx} \\
	&~~~~+2\ti{v}_3'X\log\fr{\sqx}{1+\ti{u}_4\sqx}+2\ti{b}_0''X\log X, \\
	G_3&=-4\ti{b}_1'\sqx-\ti{v}_3\bra{\fr{1}{1+\ti{u}_4\sqx}+\log\fr{\sqx}{1+\ti{u}_4\sqx}}-\ti{b}_0'\log X, \\
	G_4&=\ti{b}_0+\ti{b}_1\sqx, \\
	G_5&=0, \\
	F_4&=\fr{1}{4X^2}\bra{-\ti{b}_0+\fr{\ti{v}_4}{1+\ti{u}_4\sqx}}, \\
	F_5&=0,
	\end{split}
	\ee
where a prime here denotes $\pa/\pa\phi$.
One may further add to this any terms that vanish when the unitary gauge is chosen.
Note that, in the above expressions, we have assumed that $\phi_\mu$ is timelike because our extended cuscuton was obtained under the unitary gauge~$\phi=\phi(t)$.
If one makes a replacement~$X\to|X|$, one could incorporate a case where $\phi_\mu$ is spacelike, but this is beyond the scope of the present paper.


The case with $A_5\ne0$ can be divided into three subtypes:
(i)~$\mu_5=0$ and $\nu_5\ne0$;
(ii)~$\mu_5\ne 0$ and $\nu_5=0$;
and (iii)~$\mu_5\ne 0$ and $\nu_5\ne 0$.
One can straightforwardly
obtain the full expressions for $G_i$ and $F_j$ in each case,
but we do not present them here because the
result is too complicated to be illuminating.

\section{Propagating DOFs in non-unitary gauges}\label{app:nonU}

Throughout the main text, we work in the unitary gauge~$\phi=\phi(t)$ and specify
a class of theories where only two DOFs can propagate in this gauge.
However, one may naively think that the two-DOF nature would no longer be maintained for an inhomogeneous configuration of $\phi$.
In this appendix, we address this issue by studying the following simple example:
\begin{align}
{\cal L}=\sqrt{2|X|},\quad X\equiv -\frac{1}{2}\eta^{\mu\nu}\phi_\mu \phi_\nu,
\end{align}
which is nothing but the original cuscuton theory~\eqref{originalcuscuton} with $V(\phi)=0$ in Minkowski spacetime.
The field equation is written as
\begin{align}
2X\Box\phi +\phi^\mu \phi^\nu \phi_\mn =0,
\label{fe}
\end{align}
which is a second-order differential equation and thus there would be a propagating DOF.
Below, we show for this example that (i) if $\phi_\mu$ is timelike, $\phi$ does not propagate under a physically plausible boundary condition at spatial infinity and (ii) if $\phi_\mu$ is spacelike, the model does have a propagating DOF.
The situation here is quite similar to what happens in
the U-degenerate theory proposed in Ref.~\cite{DeFelice:2018ewo}.

\subsection{Timelike $\phi_\mu$}
Suppose $\phi$ depends only on $t$ and $x$.
If $\phi_\mu$ is timelike, we have
\begin{align}
(\phi')^2\ddot\phi -2\dot\phi\phi'\dot\phi'+\dot\phi^2\phi''=0,
\end{align}
where $\dot\phi \equiv \partial_t\phi$ and $\phi' \equiv\partial_x\phi$.
This admits the following non-unitary gauge background:
\begin{align}
\overline{\phi}=t + \alpha x,\quad  -1<\alpha<1,\quad \alpha\ne0.
\end{align}
Let us study a small fluctuation on this background:~$\phi = \overline{\phi}+\pi(t,\Vec{x})$.
The quadratic Lagrangian for $\pi$ is given by
\begin{align}
{\cal L}^{(2)}
&=-\frac{1}{2(1-\alpha^2)^{3/2}}
\left(\alpha\dot\pi-\partial_x\pi\right)^2
-\frac{1}{2(1-\alpha^2)^{1/2}}\left[
(\partial_y\pi)^2+(\partial_z\pi)^2
\right].
\end{align}
This seems to have a wrong sign kinetic term for $\alpha\neq 0$,
\begin{align}
{\cal L}^{(2)}\supset -\frac{\alpha^2}{2(1-\alpha^2)^{3/2}}\dot\pi^2,
\end{align}
implying a ghost.

The EOM for $\pi$ is given by
\begin{align}
\alpha^2\ddot\pi-2\alpha \partial_x\dot\pi +\partial_x^2\pi+
(1-\alpha^2)(\partial_y^2\pi+\partial_z^2\pi)
=0.\label{pieq}
\end{align}
Substituting $\pi=e^{-i\omega t +i\Vec{k}\cdot\Vec{x}}$, we get the following dispersion relation,
\begin{align}
(\alpha \omega -k_x)^2+(1-\alpha^2)(k_y^2+k_z^2)=0,
\end{align}
leading to the two complex solutions,
\begin{align}
\omega=\frac{k_x}{\alpha}\pm i \frac{(1-\alpha^2)^{1/2}}{\alpha}
\sqrt{k_y^2+k_z^2}.
\end{align}
Thus, apparently, one of the solutions blows up.
However, as is discussed in Ref.~\cite{DeFelice:2018ewo}, we expect that the regularity at spatial infinity removes this dangerous mode.

To see this, let us perform the following coordinate transformation:
\begin{align}
\tilde{t}=\frac{t+\alpha x}{\sqrt{1-\alpha^2}},
\quad
\tilde{x}=\frac{\alpha t+ x}{\sqrt{1-\alpha^2}},
\quad
\tilde y=y,\quad \tilde z=z.
\end{align}
Then, Eq.~(\ref{pieq}) becomes
\begin{align}
\left(\partial_{\tilde x}^2+\partial_{\tilde y}^2+\partial_{\tilde z}^2\right)\pi =0.
\end{align}
The solution to this Laplace equation which is regular at spatial infinity
is
\begin{align}
\pi = 0.
\end{align}
Therefore, the dangerous mode does not propagate if an appropriate boundary condition is imposed.

\subsection{Spacelike $\phi_\mu$}
In this case, we consider the case with $|\alpha|>1$.
We then obtain
\begin{align}
{\cal L}^{(2)}
&=-\frac{1}{2(\alpha^2-1)^{3/2}}
\left(\alpha\dot\pi-\partial_x\pi\right)^2
+\frac{1}{2(\alpha^2-1)^{1/2}}\left[
(\partial_y\pi)^2+(\partial_z\pi)^2
\right],
\end{align}
and the EOM for $\pi$ is again given by Eq.~(\ref{pieq}),
but note that now $\alpha^2>1$.
In the new coordinate system defined by
\begin{align}
\tilde{t}=\frac{\alpha t+ x}{\sqrt{\alpha^2-1}},
\quad
\tilde{x}=\frac{t+\alpha x}{\sqrt{\alpha^2-1}},
\quad
\tilde y=y,\quad \tilde z=z,
\end{align}
Eq.~(\ref{pieq}) becomes
\begin{align}
\left(-\partial_{\tilde t}^2+\partial_{\tilde y}^2+\partial_{\tilde z}^2\right)\pi=0.
\end{align}
Clearly, this is a hyperbolic equation and thus the dangerous mode
$\pi$ propagates.

%


\bibliographystyle{mybibstyle}
\bibliography{cuscuton}

\end{document}